# Possible High-Tc Superconductivity in Z = 7.33 Materials from Combinatorial Synthesis of Z = n (n=4.67, 10, 12.67) Families

O. P. Isikaku-Ironkwe[1, 2]

[1]The Center for Superconductivity Technologies (TCST)
Department of Physics,
Michael Okpara University of Agriculture, Umudike (MOUAU),
Umuahia, Abia State, Nigeria
and
[2]RTS Technologies, San Diego, CA 92122

## Abstract

The search for $MgB_2$-like superconductivity has before now been limited to binary and ternary systems. However systems with more-than-three-elements that meet the $MgB_2$ model have yet to be investigated either computationally or experimentally. Here we show that through computational synthesis of two binary or ternary materials from the Z=4.67, 10 and 12.67 families, we can get Z=7.33 systems with more-than-three elements. We have identified over 300 of such possibilities for further studies. We propose that the resultant combinatorial synthesis of such Z=7.33 systems with more-than-three-elements and that meet the $MgB_2$ model will be superconducting with Tcs higher than 39K.

## Introduction

The search for novel high-temperature superconductors has been described as "high-risk, high-payoff" project [1] that requires "passionate optimism" [2] and lots of serendipity [3, 4]. The goal is to move the field from discovery by serendipity to discovery by rational design [3, 4]. Achieving this requires understanding the chemical building blocks of superconductivity or the material specificity [5] of superconductivity. The discovery of magnesium diboride as a superconductor [6] shattered long believed search paradigms [7, 8] and introduced a new paradigm [5] of superconductivity search. It was not immediately obvious to many researchers [11, 12] that structural and electronic similarities were not sufficient conditions for basing a search for $MgB_2$-like superconductivity. This led to predictions and experimental results [9 – 12] far below the anticipated high-Tc of $MgB_2$. Turning the search to material **specific parameters in the Periodic Table such as electronegativity, valence electrons, atomic number and formula weight [5], we have been able to show that novel binary and ternary $MgB_2$-like superconductors with high-Tc do exist [13 - 21], awaiting experimental confirmation. We observe that all predicted $MgB_2$-like superconductors have been binary or ternary systems. In this paper, we extend the search to potential quaternary systems that still meet the $MgB_2$-like model. These new class of more-than-three-elements (mtte) materials is composed entirely of light elements and supersede the binaries and ternaries with similar properties proposed in [22]. These new mtte-$MgB_2$-like compounds are formed by the appropriate combinations of the Z =4.67 family [20] with either the Z =10 family [19] or the Z =12.67 family [21]. In this paper, we show the computational combinatorial synthesis process and some of the resultant compounds and their MSCDs. We also provide computational predictions of their Tcs using our Tc estimation formula [5] and superconductivity search rules.**

## Binary and Ternary Z=7.33 Superconductors

**The Z=7.33 family represents materials whose average atomic number is 7.33. $MgB_2$ is one of them. Many binary and ternary Z=7.33 materials have been computationally proposed that show $MgB_2$-like superconductivity [13 - 21]. Table 1 shows some of them and their MSCDs. Their Tcs have been estimated using the new material specific Tc equation derived in [5], given by:**

$$T_c = \mathcal{X} \frac{\mathrm{Ne}}{\sqrt{Z}} K_o \quad (1)$$

**where Ko = n(Fw/Z) and n can be estimated empirically. For magnesium diboride, n = 3.65. Fw is formula weight and Z is average atomic number. The value of n may be less for other Z =7.33 materials.**

## Combinatorial Synthesis of Novel Z=7.33 Materials

**There have been no reports of the existence of more-than-three-elements (mtte)-$MgB_2$-like superconductors. Such materials would be required to have the same valence electron count** and atomic number as $MgB_2$. We propose that mtte-$MgB_2$-like superconductors could result from the combinatorial synthesis of two families (Fz): where Z=4.67, 10 or 12.67 materials, with Ne=2.67 in each case, according to the equations:

$$F_{Z=4.67} + F_{Z=10} = F_{Z=7.33} \quad (2)$$

$$2F_{Z=4.67} + F_{Z=12.67} = F_{Z=7.33} \quad (3)$$

These combinations are illustrated in figures 1 and 2. Tables 2, 3 and 4 show some of the potential materials in the three families under consideration. Equation (2) states that 1:1 combinatorial synthesis of the $Fz_{=4.67}$ and $Fz_{=10}$ families will yield a new family $Fz_{=7.33.}$ From equation (3), the combinatorial synthesis of two parts of a $Fz_{=4.67}$ family member with one part of a $Fz_{=12.67}$ family member, will be a new compound belonging to the family $Fz_{=7.33}$. The expected MSCD [5] of the new compound will be:

$$\text{MSCD of compound} = \langle \mathcal{X}, \text{Ne}, \text{Z}, \text{Ne}/\sqrt{Z}, \text{Fw}, \text{Fw/Z} \rangle \quad (4)$$
$$= \langle \mathcal{X}, 2.667, 7.333, 0.9847, \text{Fw}, \text{Fw/Z} \rangle$$

where Ne is the valence electron count of each family = 2.667, the average atomic number Z =7.333 and the average final electronegativity, $\mathcal{X}$. Fw is the total formula weight of the new Z=7.33 material. In [5] we showed that a high-Tc should be expected when:

$$0.8 < \text{Ne}/\sqrt{Z} < 1.0 \quad (5)$$

Thus from equation (4) and (5) the resultant combinatorial synthesis of the families as defined in equations (2) and (3), should lead to $MgB_2$-like materials with high Tc.

## Combinatorial Results

Tables 5 and 6 show some of the results of implementing the combinatorial synthesis of two families ($F_{Z=4.67}$ and $F_{Z=10}$ ) of materials with the same Ne =2.67 in the 1:1 ratio indicated. In each case we end up with $MgB_2$-like materials with Z=7.333 and Ne=2.67 as predicted from equation (4) above. The number of possible compounds is the product of the numbers of compounds in each family, in this case 170. In Table 6, we show just 12 out of the 170 MSCDs possibilities. As observed, some will be 3-, 4-, 5- or 6-elements $MgB_2$-like materials with Z=7.33 while the electronegativities vary. From Table 6, except for the 3-atom case with Fw/Z

**=6.26, the 6-atom cases have average Fw/Z of 12.52. Tables 7 and 8 show some of the results of computational combinatorial synthesis of two parts of a member of family $F_{Z=4.67}$ with one part of a member of the $F_{Z=12.67}$ family. With 10 members of $F_{Z=4.67}$ and 16 members in $F_{Z=12.67}$, we have 160 minimum possible combinations. Again, some will be 3-, 4-, 5- or 6-elements $MgB_2$-like materials with Z=7.33 while the electronegativity varies. Only 6 possible results are listed in Table 7 with their MSCDs in Table 8. From Table 8 we note that we get 9-atom systems with 4 or 5 elements. The average Fw/z is 18.77. We summarize the variation of Fw/Z with number of atoms in Table 9 and plot Fw/z vs computed Tc in Figure 3.**

## Analysis of Combinatorial Results

**Figures 1 and 2 shows the combinatorial process which results in the creation of Z=7.33 family, noted for hosting $MgB_2$. A significant aspect of this process is the creation of 4-, 5-, 6-element compounds that meet the model of $MgB_2$. If they are found to exist and are stable, we may have for the first time, materials with Tcs higher than $MgB_2$ but meet its specifications. This will be because Fw/z is higher as explained in [5] with the derived equation for Tc [5] in terms of electronegativity, $\mathcal{X}$, valence electron count, Ne, atomic number, Z, and formula weight, Fw.**

## Discussion

**Rosner et al. [9] proposed that hole-doping LiBC, a Z=4.67 material to $Li_{0.5}BC$ might lead to higher Tc than found in $MgB_2$. Though their concept was right, the approach was not. Working with the paradigm that such doping should not change the valence electron count and the average atomic number, we have been able to "dope" Z=4.67 materials with Z = 10.0 and Z=12.67 materials resulting in Z=7.33 materials with Fw/Z much higher than that of $MgB_2$. Looking at the elements involved we see that they are mostly combinations of the 12 lightest elements with Z<20 in such a manner as to give an average Z= 7.333 and Ne=2.667. The electronegativity rarely exceeds 1.733. For Fw/Z of 12.52 the computed Tc is 63K for the 6-** **atom systems with Z=7.33 MSCD and electronegativity same as $MgB_2$. The highest Fw/Z for these compounds was 18.77 which we have computed should correspond to a Tc of 75K.**

## Conclusions

**This research revealed over 300 potential new high-Tc $MgB_2$-like superconductors in the hitherto unexplored chemical world of 6- and 9- atom mtte-compounds with average atomic number, Z, of 7.333 and average valence electron count, Ne, of 2.667. These novel 6-atom and 9-atom $MgB_2$-like materials with Fw/Z as high as 18.77 strongly suggest that they may have Tcs as high as 75K (table 9 and figure 3). Experimental confirmation of these predictions presents a strong challenge to experimentalists and computational material scientists and opens a whole new world of mtte-$MgB_2$-like superconductors, too early to evaluate.**

## Acknowledgements

**The author acknowledges useful discussions with Professor M. Brain Maple at U.C. San Diego. This research was executed with grants from Dr. M. J. Schaffer, formerly at General Atomics, San Diego. Research literature from Dr. Jim O'Brien facilitated my work.**

## Tables

| Material | | $\mathcal{X}$ | Ne | Z | $Ne/\sqrt{Z}$ | Fw | Fw/Z |
|---|---|---|---|---|---|---|---|
| 1 | $MgB_2$ | 1.733 | 2.6667 | 7.333 | 0.9847 | 45.93 | 6.263 |
| 2 | $Li_2S$ | 1.5 | 2.6667 | 7.333 | 0.9847 | 45.95 | 6.266 |
| 3 | $Be_2Si$ | 1.6 | 2.6667 | 7.333 | 0.9847 | 46.11 | 6.290 |
| 4 | LiBeP | 1.533 | 2.6667 | 7.333 | 0.9847 | 46.92 | 6.398 |
| 5 | LiMgN | 1.733 | 2.6667 | 7.333 | 0.9847 | 45.26 | 5.999 |
| 6 | MgBeC | 1.733 | 2.6667 | 7.333 | 0.9847 | 45.33 | 6.181 |
| 7 | LIBSi | 1.6 | 2.6667 | 7.333 | 0.9847 | 45.84 | 6.251 |

**Table 1** MSCD of some 7 binary and ternary Z =7.333 materials, with Ne=2.667. $\mathcal{X}$ = electronegativity, Ne = valence electron count, Z= average atomic number and Fw = formula weight Reference [5] describes how to compute MSCD.

| Material | | $\mathcal{X}$ | Ne | Z | $Ne/\sqrt{Z}$ | Fw | Fw/Z |
|---|---|---|---|---|---|---|---|
| 1 | LiBC | 1.8333 | 2.6667 | 4.6667 | 1.2344 | 29.76 | 6.377 |
| 2 | $BeB_2$ | 1.8333 | 2.6667 | 4.6667 | 1.2344 | 30.63 | 6.564 |
| 3 | $Be_2C$ | 1.8333 | 2.6667 | 4.6667 | 1.2344 | 30.03 | 6.435 |
| 4 | $Li_2O$ | 1.8333 | 2.6667 | 4.6667 | 1.2344 | 29.88 | 6.403 |
| 5 | LiBeN | 1.8333 | 2.6667 | 4.6667 | 1.2344 | 29.96 | 6.420 |
| 6 | $LiB_5$ | 1.8333 | 2.6667 | 4.6667 | 1.2344 | 60.99 | 13.07 |
| 7 | $Li_3BN_2$ | 1.8333 | 2.6667 | 4.6667 | 1.2344 | 59.65 | 12.78 |
| 8 | $NaAlH_4$ | 1.8 | 1.3333 | 4.6667 | 0.6172 | 54.01 | 11.574 |
| 9 | $KBH_4$ | 1.8667 | 1.3333 | 4.6667 | 0.6172 | 53.95 | 11.56 |
| 10 | $MgH_2$ | 1.8 | 1.3333 | 4.6667 | 0.6172 | 26.33 | 5.642 |

**Table 2:** 10 MSCDs of some Z =4.67 materials with Ne=2.67 and 1.33

| Material | | $\mathcal{X}$ | Ne | Z | Ne/$\sqrt{Z}$ | Fw | Fw/Z |
|---|---|---|---|---|---|---|---|
| 1 | MgBeSi | 1.833 | 2.667 | 10.0 | 0.8433 | 61.41 | 6.141 |
| 2 | $Na_2O$ | 1.767 | 2.667 | 10.0 | 0.8433 | 61.98 | 6.198 |
| 3 | KBeN | 1.767 | 2.667 | 10.0 | 0.8433 | 62.12 | 6.212 |
| 4 | NaMgN | 1.7 | 2.667 | 10.0 | 0.8433 | 61.31 | 6.131 |
| 5 | LiCaN | 1.667 | 2.667 | 10.0 | 0.8433 | 61.03 | 6.103 |
| 6 | $CaB_2$ | 1.667 | 2.667 | 10.0 | 0.8433 | 61.70 | 6.170 |
| 7 | CaBeC | 1.667 | 2.667 | 10.0 | 0.8433 | 61.10 | 6.110 |
| 8 | $Mg_2C$ | 1.633 | 2.667 | 10.0 | 0.8433 | 60.63 | 6.063 |
| 9 | NaAlC | 1.633 | 2.667 | 10.0 | 0.8433 | 61.98 | 6.198 |
| 10 | BeBSc | 1.6 | 2.667 | 10.0 | 0.8433 | 64.78 | 6.478 |
| 11 | NaBSi | 1.567 | 2.667 | 10.0 | 0.8433 | 61.89 | 6.189 |
| 12 | $BeAl_2$ | 1.5 | 2.667 | 10.0 | 0.8433 | 62.97 | 6.297 |
| 13 | $Be_2Ti$ | 1.5 | 2.667 | 10.0 | 0.8433 | 65.90 | 6.590 |
| 14 | NaBeP | 1.5 | 2.667 | 10.0 | 0.8433 | 62.97 | 6.297 |
| 15 | LiNaS | 1.467 | 2.667 | 10.0 | 0.8433 | 62.00 | 6.200 |
| 16 | LiMgP | 1.433 | 2.667 | 10.0 | 0.8433 | 62.22 | 6.222 |
| 17 | LiAlSi | 1.433 | 2.667 | 10.0 | 0.8433 | 62.01 | 6.201 |

**Table 3:** 17 MSCDs of Z =10.0 materials.

| Material | | $\mathcal{X}$ | Ne | Z | $Ne/\sqrt{Z}$ | Fw | Fw/Z |
|---|---|---|---|---|---|---|---|
| 1 | CaBeSi | 1.433 | 2.667 | 12.667 | 0.7493 | 77.18 | 6.09 |
| 2 | $Na_2S$ | 1.433 | 2.667 | 12.667 | 0.7493 | 78.05 | 6.16 |
| 3 | LiKS | 1.433 | 2.667 | 12.667 | 0.7493 | 78.11 | 6.17 |
| 4 | NaAlSi | 1.4 | 2.667 | 12.667 | 0.7493 | 78.06 | 6.16 |
| 5 | $Mg_2Si$ | 1.4 | 2.667 | 12.667 | 0.7493 | 76.71 | 6.06 |
| 6 | NaMgP | 1.4 | 2.667 | 12.667 | 0.7493 | 78.27 | 6.18 |
| 7 | KAlC | 1.6 | 2.667 | 12.667 | 0.7493 | 78.09 | 6.16 |
| 8 | NaCaN | 1.633 | 2.667 | 12.667 | 0.7493 | 77.08 | 6.09 |
| 9 | $LiBGe_{0.89}Si_{0.11}$ | 1.6 | 2.667 | 12.667 | 0.7493 | 85.46 | 6.75 |
| 10 | KBSi | 1.533 | 2.667 | 12.667 | 0.7493 | 78.0 | 6.16 |
| 11 | KBeP | 1.467 | 2.667 | 12.667 | 0.7493 | 79.09 | 6.24 |
| 12 | $MgAl_2$ | 1.4 | 2.667 | 12.667 | 0.7493 | 78.27 | 6.18 |
| 13 | LiAlTi | 1.333 | 2.667 | 12.667 | 0.7493 | 81.8 | 6.46 |
| 14 | LiCaP | 1.367 | 2.667 | 12.667 | 0.7493 | 77.99 | 6.16 |
| 15 | $Na_3AlP_2$ | 1.4 | 2.667 | 12.667 | 0.7493 | 157.89 | 12.465 |
| 16 | $K_3BN_2$ | 1.7333 | 2.667 | 12.667 | 0.7493 | 156.13 | 12.326 |

**Table 4:** 16 MSCDs of Z =12.667 materials.

| | Fz=4.67 | Fz=10 | 12 of the combinatorial results: Fz=4.67 + Fz=10 for 170 combinatorial possibilities. | |
|---|---|---|---|---|
| 1 | $BeB_2$ | $Na_2O$ | $Na_2BeB_2O$ | LiNaBeBCP |
| 2 | $Li_2O$ | $Mg_2C$ | $LiNa_2BCO$ | $LiNa_2B_5O$ |
| 3 | $Be_2C$ | NaBeP | $Li_3MgPO$ | $Li_2NaAlCO$ |
| 4 | $MgH_2$ | KBeN | $Na_2Be_2CO$ | $Li_2Mg_2CO$ |
| 5 | $LiB_5$ | LiMgP | MgBeC | $NaBe_2AlC_2$ |
| 6 | LiBC | NaAlC | $LiMg_2BC_2$ | $LiNaBAlC_2$ |

**Table 5: 12 of the combinatorial results for 170 combinatorial possibilities from the two families of materials (Fz=4.67 + Fz=10).**

| Material | | $\mathcal{X}$ | Ne | Z | $Ne/\sqrt{Z}$ | Fw | Fw/Z |
|---|---|---|---|---|---|---|---|
| 1 | $Na_2BeB_2O$ | 1 .8 | 2.6667 | 7.3333 | 0.9847 | 92.61 | 12.63 |
| 2 | $LiNa_2BCO$ | 1.8 | 2.6667 | 7.3333 | 0.9847 | 91.74 | 12.51 |
| 3 | $Li_3MgPO$ | 1.6333 | 2.6667 | 7.3333 | 0.9847 | 92.10 | 12.56 |
| 4 | $Na_2Be_2CO$ | 1.8 | 2.6667 | 7.3333 | 0.9847 | 92.01 | 12.55 |
| 5 | MgBeC | 1.7333 | 2.6667 | 7.3333 | 0.9847 | 45.93 | 6.26 |
| 6 | $LiMg_2BC_2$ | 1.7333 | 2.6667 | 7.3333 | 0.9847 | 90.39 | 12.34 |
| 7 | $Li_2Mg_2CO$ | 1.7333 | 2.6667 | 7.3333 | 0.9847 | 90.51 | 12.34 |
| 8 | $Li_2NaAlCO$ | 1.7333 | 2.6667 | 7.3333 | 0.9847 | 91.86 | 12.53 |
| 9 | LiNaBeBCP | 1.6667 | 2.6667 | 7.3333 | 0.9847 | 92.73 | 12.64 |
| 10 | $Mg_3H_2C$ | 1.7167 | 2. 0 | 7.3333 | 0.7386 | 86.96 | 11.86 |
| 11 | $NaBe_2AlC_2$ | 1.7333 | 2.6667 | 7.3333 | 0.9847 | 92.01 | 12.55 |
| 12 | $LiNaBAlC_2$ | 1.7333 | 2.6667 | 7.3333 | 0.9847 | 91.74 | 12.51 |

**Table 6:** **MSCDs of 12 out of 170 combinatorial materials possibilities from (Fz=4.67 + Fz=10). The combinations result in 6-atom compounds with 3, 4, 5 and 6 different elements.**

| | $F_{z=4.67}$ | $F_{z=12.667}$ | 6 of the combinatorial results: $2F_{z=4.67} + F_{z=12.67}$ for 160 combinatorial possibilities. |
|---|---|---|---|
| 1 | $Li_2O$ | $Mg_2Si$ | $Li_4Mg_2SiO_2$ |
| 2 | $Be_2C$ | $Na_2S$ | $Na_2Be_4C_2S$ |
| 3 | LiBC | CaBeSi | $CaBe_5SiC_2$ |
| 4 | LiBeN | KBSi | $KBe_4BSiC_2$ |
| 5 | $LiB_5$ | LiAlTi | $LiBe_4AlTiC_2$ |
| 6 | $Li_3BN_2$ | LiCaP | $Li_2Na_2B_2C_2S$ |

**Table 7:** 6 of the combinatorial results for 160 combinatorial possibilities in ($2F_{z=4.67} + F_{z=12.67}$).

| | Material | $\mathcal{X}$ | Ne | Z | Ne/$\sqrt{Z}$ | Fw | Fw/Z |
|---|---|---|---|---|---|---|---|
| 1 | $Li_4Mg_2SiO_2$ | 1.689 | 2.6667 | 7.3333 | 0.9847 | 136.47 | 18.61 |
| 2 | $Na_2Be_4C_2S$ | 1.7 | 2.6667 | 7.3333 | 0.9847 | 138.11 | 18.33 |
| 3 | $CaBe_5SiC_2$ | 1.7 | 2.6667 | 7.3333 | 0.9847 | 137.24 | 18.72 |
| 4 | $KBe_4BSiC_2$ | 1.733 | 2.6667 | 7.3333 | 0.9847 | 138.06 | 18.83 |
| 5 | $LiBe_4AlTiC_2$ | 1.667 | 2.6667 | 7.3333 | 0.9847 | 141.86 | 19.35 |
| 6 | $Li_2Na_2B_2C_2S$ | 1.7 | 2.6667 | 7.3333 | 0.9847 | 137.57 | 18.76 |

**Table 8:** MSCD of 6 out of 160 combinatorial materials possibilities from ($F_{z=4.67} + F_{z=10}$) of data from tables 2, 4 and 7. The combinations result in 9-atom compounds with 4 and 5 elements.

| #Atoms (An) | Fw/Z(Ave) | n | Ko=n(Fw/Z) | Tc(K) | Example | Reference |
|---|---|---|---|---|---|---|
| 3 | 6.26 | 3.65 | 22.85 | 39 | $MgB_2$ | Figure 2 of [5] |
| 6 | 12.52 | 2.95 | 36.91 | 63 | $NaBe_2AlC_2$ | Table 6 of this paper + Figure 2 of [5] |
| 9 | 18.77 | 2.34 | 43.94 | 75 | $KBe_4BSiC_2$ | Table 8 of this paper + Figure 2 of [5] |

**Table 9: Data on variation of Fw/Z with number of atoms for the Z=7.33 family of materials of which $MgB_2$ belongs. All these examples have same $\mathcal{X}$, Ne, and Z , but different Fw/Z. Using the material specific Tc formula: $\mathbf{T_c} = \mathcal{X}\frac{\mathrm{Ne}}{\sqrt{Z}}K_o$ derived in reference [5].**

## Figures

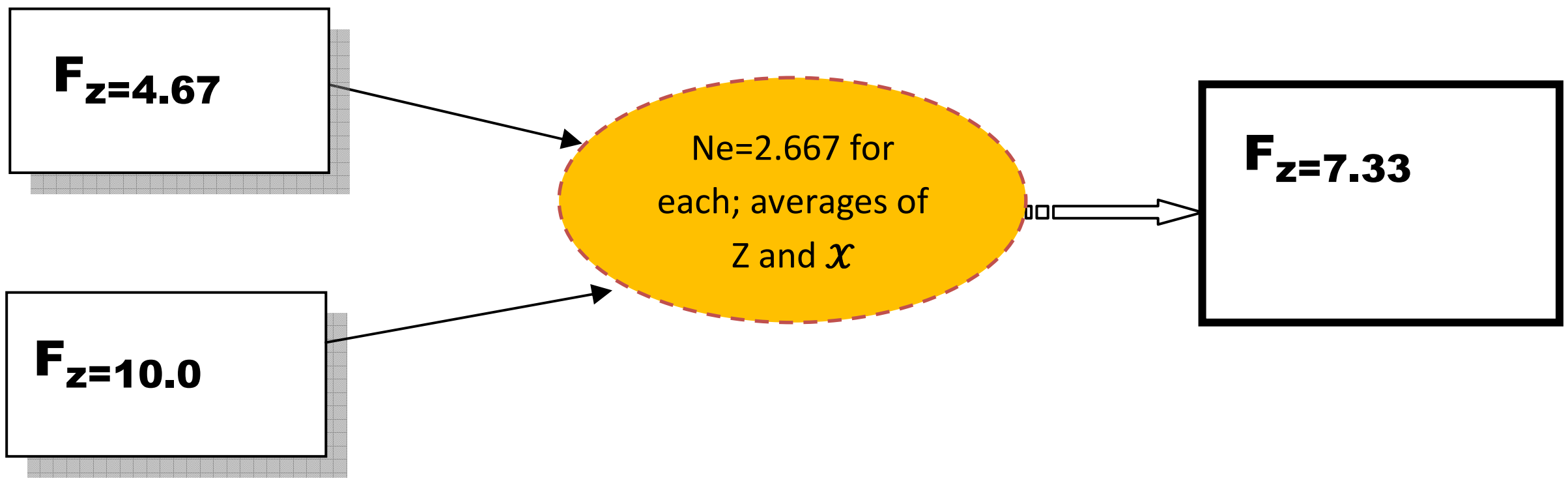

**Figure 1**: **Combining materials from Fz=4.67 and Fz=10.0 yields a new family Fz=7.33 material with Ne=2.67 and Z=7.33. The electronegativity will be an average of the electronegativities of Fz=4.67 and Fz=10.0**

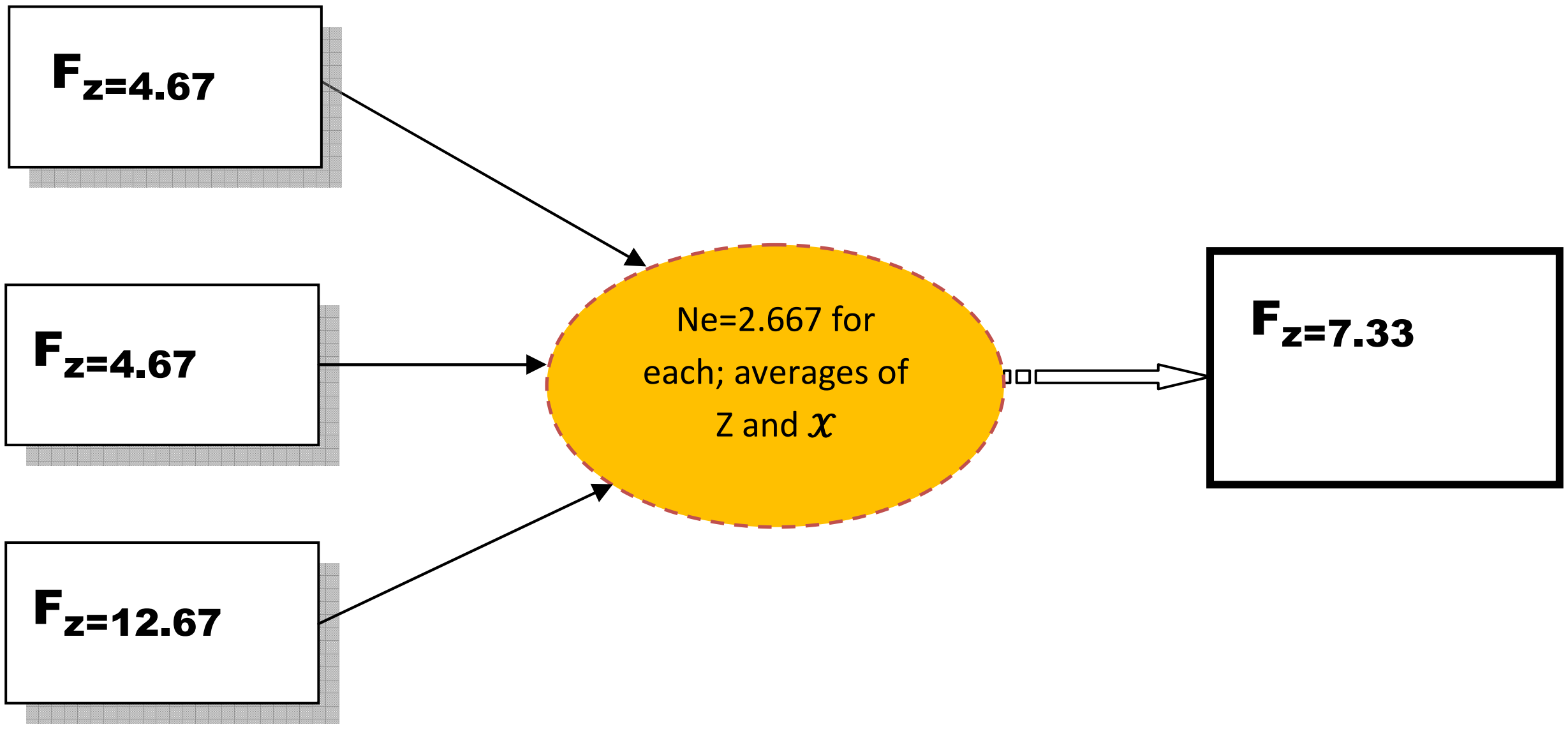

**Figure 2**: **Combining 2x a material from Fz=4.67 with one material from Fz=12.67 yields a new family Fz=7.33 material with Ne=2.67 and Z=7.33. The electronegativity will be an average of the electronegativities of 2(Fz=4.67) and Fz=12.67.**

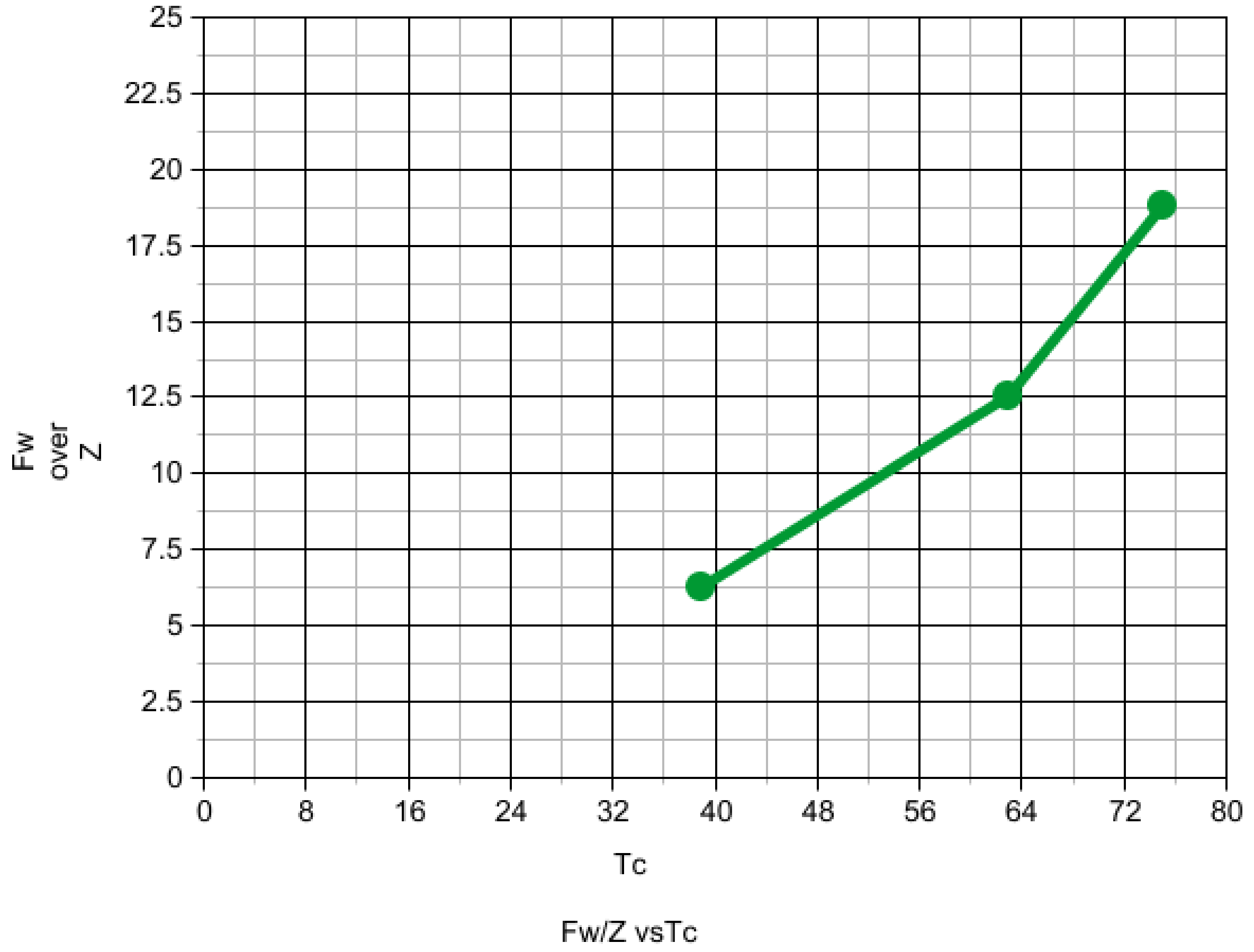

**Figure 3:** Fw/Z versus Tc. From Reference[5] we know that Tc increases with Fw/Z. The Z=7.33 materials with the same MSCD as MgB2 should have the same Tc but since they have higher Fw/Z, we should expect a higher Tc by equation (1) in this paper and figure 2 of Ref [5].